# The Triple Helix, Quadruple Helix, …, and an *N*-tuple of Helices: Explanatory Models for Analyzing the Knowledge-based Economy?

Loet Leydesdorff

Amsterdam School of Communication Research (ASCoR), University of Amsterdam

Kloveniersburgwal 48, 1012 CX Amsterdam, The Netherlands

loet@leydesdorff.net ; http://www.leydesdorff.net

**Abstract**

Using the Triple Helix model of university-industry-government relations, one can measure the extent to which innovation has become systemic instead of assuming the existence of national (or regional) systems of innovations on *a priori* grounds. Systemness of innovation patterns, however, can be expected to remain in transition because of integrating and differentiating forces. Integration among the functions of wealth creation, knowledge production, and normative control takes place at the interfaces in organizations, while exchanges on the market, scholarly communication in knowledge production, and political discourse tend to differentiate globally. The neo-institutional and the neo-evolutionary versions of the Triple Helix model enable us to capture this tension reflexively. Empirical studies inform us whether more than three helices are needed for the explanation. The Triple Helix indicator can be extended algorithmically, for example, with local-global as a fourth dimension or, more generally, to an *N*-tuple of helices.

**Introduction**

Etzkowitz & Leydesdorff (1995 and 2000) proposed the model of a Triple Helix of university-industry-government relations for explaining structural developments in knowledge-based economies. In a knowledge-based economy—as against a political economy—the structure of society is continuously upset by transformations which originate from the techno-sciences. The relevant framework of society can thus be expected to have changed. Using the Triple Helix (TH) model, analysis can be more specific than by claiming a generalized transition from "mode 1" to "mode 2" in "the new production of knowledge" (Gibbons *et al.*, 1994). While in a political economy only two types of communication are prevalent—(*i*) the equilibrium-seeking dynamics of markets and (*ii*) normative control mechanisms along the public-private interface—a third subdynamic, namely (*iii*) the equilibrium-upsetting dynamics of socially organized knowledge production, has also to be considered in the analysis of knowledge-based economies.

This model improves on the non-linear model that replaced linear models based on "market pull" or "technology push." The chaining model of Kline & Rosenberg (2006) and the model of trajectories and regimes (Dosi, 1982; Nelson & Winter, 1982) were developed mainly at the level of the firm (Casson, 1997). At the level of society, however, the theory of "national systems of innovation" has been dominant (Lundvall, 1988 and 1992; Nelson, 1993; cf. Braczyk *et al.*, 1998). The TH model improves on this "(national) systems of innovation" model because it no longer requires the assumption *ex ante*, for example, of "national" or "regional" systems for the integration. One can formulate the empirical question of whether a system has emerged at the national or regional level (Leydesdorff & Zawdie, 2010). Emerging systemness can then be analyzed in terms of potential synergies among three subdynamics (or perhaps more; cf. Carayannis & Campbell, 2009 and 2010; Marcivich & Shinn, 2010). Can the non-linear interactions among an *N*-tuple of helices be specified and measured? Under what circumstances can increased interactions be expected to lead to synergies? If so, at what level can synergies be retained, by whom, and in what respects?

In a series of case studies of national systems, for example, we could show that the Netherlands can be considered as a *national* system of innovations (given a specific operationalization of the three subdynamics in terms of indicators; Leydesdorff *et al*., 2006), but Germany cannot because synergies at the level of the federal states (e.g., Bavaria) prevail (Leydesdorff & Fritsch, 2006). In the case of Hungary, Lengyel & Leydesdorff (in press) have found that in this country three *regional innovation systems* have replaced the centrally coordinated national system following the transition in the early 1990s and the gradual accession to the EU in the period thereafter: the capital (Budapest) can nowadays be characterized as a metropolitan innovation system which competes with other such cities (Vienna, Munich); the western part of the country has been firmly integrated into a Europeanized innovation system because of foreign direct investments; and the eastern part of the country has remained a state-led innovation system. Each of these three regions has its own specific form of integration and differentiation, which can be expected to result in different synergies.

Similarly, one can analyze whether innovation systems are technology-specific or sector-based (Pavitt, 1984; Carlsson, 2006). These are empirical questions which Triple Helix indicators enable us to address. Using co-authorship data from the *Science Citation Index*, for example, Leydesdorff & Sun (2009) showed that in Japan university-industry-government relations—and particularly university-industry collaborations—have declined during the last two decades despite explicit policies to stimulate such relations and the proclaimed success of these policies. University scholars in Japan have increasingly coauthored with foreign colleagues, thus favouring internationalization above industrial relevance. Similar, but also somewhat different developments were found in Korea (Park *et al*., 2010). In Japan, however, including internationalization as a fourth dimension in the design improved the explanation.

In other words, policy analysts cannot conclude from the model or the results of these studies that university-industry-government relations have been intensified and are still increasing (Etzkowitz, 2008; Hessels & Van Lente, 2008). The relevant systems have been changed and continue to evolve since the disturbances of equilibrium by knowledge-based transformations are understood no longer as a consequence of "creative destruction" by entrepreneurs only (Nelson & Winter, 1982; Schumpeter, 1912, [1939] 1964), but as a structural characteristic of knowledge-based economies (Cooke, 2002; Dasgupta & David, 1984; Foray, 2004; Leydesdorff, 2006). The knowledge-based subdynamics of the social system tends to dissolve and reconstruct the "systemness" of previous states by restructuring (natural) resources into epistemic objects and thus changing the expectations of stakeholders about possible and desirable relations (Leydesdorff, 2010a, and in preparation). For example, the Netherlands has become a major exporter of tomatoes although tomatoes cannot grow naturally given the climate in this country. In a knowledge-based system, nothing is given "naturally" which cannot be deconstructed analytically, improved upon, and innovatively reconstructed.

In institutional terms, this structural change from a political to a knowledge-based economy has made the universities most salient to the system (Godin & Gingras, 2000; Halffman & Leydesdorff, 2010). In other words, systems can no longer be presumed to "exist" like "national states" or their governments. The expectation of emerging systemness can continuously be tested against the available data, and higher education and knowledge-based updates are needed (Pasinetti, 1993). If a system or systemness is indicated, it can be expected to remain in transition (Carayannis & Campbell, 2006; Etzkowitz & Leydesdorff, 1998). The construction of specific advantages and the retention of wealth from these advantages is a policy objective more relevant than, and different from, the sustainability of an assumed system (Cooke & Leydesdorff, 2006; cf. Geels & Schot, 2007).

**From two to three subdynamics**

Let me shortly recapitulate the history and origins of the Triple Helix model in order to explain the difference between the analytical model for explaining knowledge-based socio-economic developments and the metaphor of stimulating university-industry-government relations proclaimed by state agencies in political discourse (e.g., the Swedish Vinnova). These policy models can be developed further into a program for industrial stimulation which takes the knowledge factor into account (Etzkowitz, 2005; Mirowski & Sent, 2007). However, I distinguish between this neo-institutional approach and the research program about (e.g., national or regional) networks of relations between universities, industries, and governments, on the one hand, and on the other, a neo-evolutionary research program about possible synergies among functions such as wealth creation, knowledge production, and normative control. While institutions and inter-institutional arrangements can be stimulated by local or national governments, markets and sciences operate at the global level. From this neo-evolutionary perspective, the function of institutional agency thus involves distributed instances. The distributions of observable occurrences can be tested statistically for their significance against the expectations.

The Triple Helix has attempted to bring these two intellectual perspectives together (Leydesdorff & Zawdie, 2010). At a workshop which I organized in Amsterdam (with Peter van den Besselaar) in 1993 about "Evolutionary Economics: New Directions in Technology Studies" (Leydesdorff & Van den Besselaar, 1994), Henry Etzkowitz contributed with a paper entitled "Academic-Industry Relations: A Sociological Paradigm for Economic Development" (Etzkowitz, 1994). In an epilogue to the edited volume that came out of this workshop, I summarized the results using the depiction of a hypercycle as reproduced in Figure 1. I argued that only two—instead of three—dynamics were postulated in Schumpeter's (1939) model of innovation: factor substitution in terms of changes along the production function versus technological developments as changes of the production function towards the origin (Sahal, 1981).

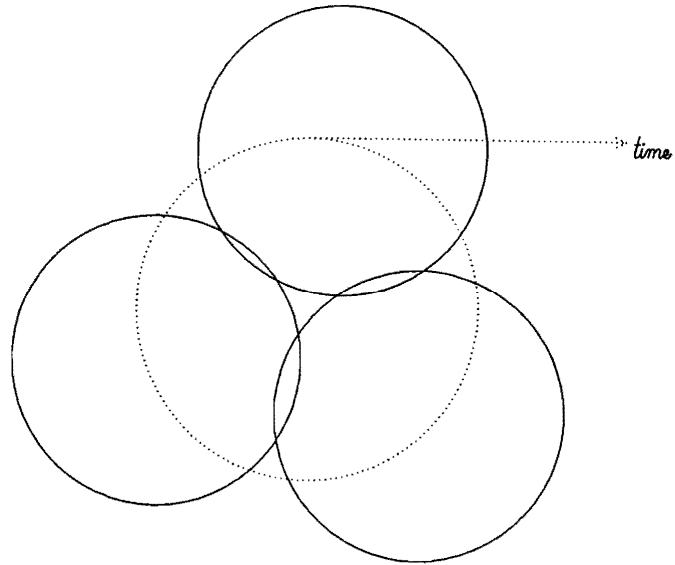

**Figure 1**: The hypercycle model as depicted in Leydesdorff (1994, at p. 186).

When knowledge-based innovations are considered, more than two—i.e., at least three—subdynamics have to be declared in the model (Nelson & Winter, 1982). Figure 1 shows that the three spheres (or helices when extended over time) do not need to be coordinated into a central overlapping zone (although they might be). As I shall suggest below, this failing overlap can be characterized as an overlap with a minus sign. This differentiation—as against (positive) integration in a central zone—offers the possibility of another synergetic mechanism, namely selective adjustment of the cycles to one another over time (Carayannis & Ziemnowicz, 2007).

Such a synergy presumes the operation of three helices as selection mechanisms upon one another. From the perspective of evolution theory, a single (e.g., "natural") selection mechanism leads to an evolutionary model; two selection mechanisms can lead to a co-evolution; three selection mechanisms allow not only for the selection of certain selections for stabilization (e.g., along a trajectory in a coevolution), but also for the selection of some stabilizations for globalization, that is, at the regime level. When more than two helices are involved, all bets are off, since various kinds of chaotic behaviour become possible (Leydesdorff & Van den Besselaar, 1998; Li & Yorke, 1975). A stabilization along a trajectory can be hyper-stabilized (as in the case of "lock in") or destabilized, meta-stabilized, and globalized into a regime by interacting with a third selection mechanism (Dolfsma & Leydesdorff, 2009).

The helices operate as selection mechanisms asymmetrically on one another, but mutual selections may shape a trajectory as in a co-evolution. However, the hypercycle operates at a next-order regime level which is global with respect to the (three or more) underlying subdynamics. Consequently, its effects cannot be reduced to the contributions of specific subdynamics because of the expectation of nonlinear interactions in the loops. The cybernetic principle is that construction of this overlay is bottom-up, but control emerges thereafter top-down (as feedback). In addition to interactions between each two (in potential co-evolutions), the bi-lateral interactions can also begin to interact among themselves and lead to three-way interaction effects that can function positively or negatively as feedforward or feedback mechanisms, respectively.

At the level of society, these dynamics of exchanges and communication in different domains are structured by mutual expectations which limit the scope of the possible dynamics. This structuration of expectations into an overlay is structurally different from the perspectives that can be generated within the various interacting helices in isolation. The relations are asymmetrical selections and interact, so that their effects at the overlay level can be unintended. Remember that selection is a deterministic process; the specification of the system of reference thus matters. Has a regime been operating or a trajectory? If both, then the next question is obviously: "to what extent?" While the dynamics within each of the helices develop primarily along their own internal axes using specific codes of communication, the dynamics at the overlay level are continuously disturbed by new developments within each of the helices on which the overlay depends.

The knowledge-based subdynamics operate more than the other two (of price mechanism and normative control) in terms of shaping and reconstructing expectations. Since the overlay operates in terms of mutual expectations, the knowledge-intensity drives the knowledge-based system into a next gear in which material conditions can be traded off against and gradually replaced by perspectives of technological change. New states and sets of new possible states of a system can be envisaged in the present, and thus redundancy is continuously increased (Leydesdorff, 2010b). In other words, cultural evolution counteracts the biological evolution from which it has emerged and on which it rests. For example, the polder vegetation in my garden has replaced the ecology in the lake which was there in times past.

In 1993-1994 I lacked the metaphors needed to express this theoretical perspective in terms with sufficient policy relevance. Fortunately, Henry Etzkowitz and I met again in the following year at a workshop in Sweden, and Henry suggested a new project based on a future collaboration. I asked him "about what?" and when he replied "university-industry relations," I answered: "Two is not complex enough for me; I need at least three subdynamics!" From these discussions, the metaphor of a Triple Helix of university-industry-government relations emerged in the months thereafter when we began to prepare for the first Triple Helix meeting in 1996 in Amsterdam (Etzkowitz & Leydesdorff, 1995).[1]

The tension between our two models—Etzkowitz' neo-institutional one focusing on inter-institutional networking and exchanges, and my model of neo-evolutionary mechanisms of exchanges among functions (wealth creation, knowledge production, and normative control)—provided heuristics for a number of years. After the articles in the early 2000s, however, we decided to stop coauthoring because of possible misunderstandings and confusions about the two models and their graphical representations. The hypercycle model depicted above (in Figure 1) is different from—almost the opposite of—Figure 3 at p. 112 in Etzkowitz & Leydesdorff (2000), which depicts a (hierarchical) model of institutional university-industry-government *relations*. Relational models are by definition hierarchical (that is, based on relations among relations), whereas functions operate in terms of *positions* of latent variables in the vector space that is first spanned—as in an architecture—in terms of the relations among the variables.

---

[1] We agreed to use this metaphor in our email exchanges of November 1994. Peter Healey informed me in March 2004 that he had used this metaphor at a meeting in Mexico in January 1993, but never published it. Note that a Triple Helix model was the (erroneous) alternative of Linus Pauling for explaining the structure of DNA as against the Double Helix proposed by James D. Watson and Francis Crick in 1953; the latter two scholars received the Nobel Prize in 1962 (Leydesdorff & Etzkowitz, 2003; cf. Lewontin, 2000).

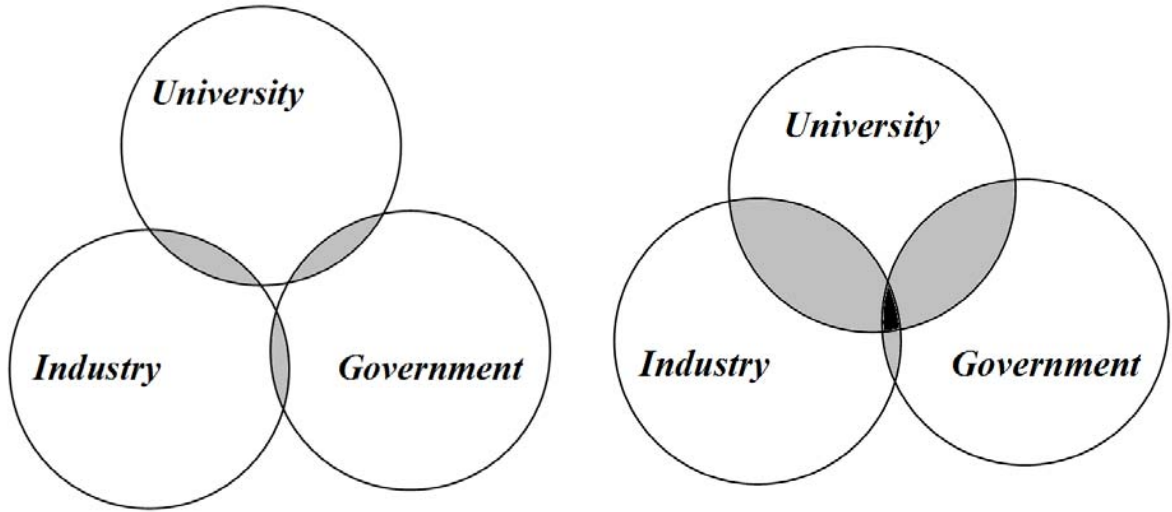

**Figure 2**: A Triple Helix configuration with negative and positive overlap among the three subsystems.

Alternatively, in my opinion, one can oppose the two models as in Figure 2. In the left-hand panel, an overlay can be penciled on top of the missing (and therefore *negative*) overlap (as in Figure 1). This hypercycle can be interpretated as a decentralized process of translations that can theoretically be appreciated as an overlay of expectations in a knowledge-based economy. However, the interacting translations can be expected to shape this overlay to a variable extent. Note that the translations are among communications of a different nature (industrial, academic, and political) or—in the terminology of Parsons (1968) and Luhmann (1995, 1997)—of functionally differentiated codes of communication. The translations operate in terms of functions (wealth creation, etc.) of communications that are analytically different from institutions or agents (Leydesdorff & Zawdie, 2010).

**The fourth helix and beyond**

The metaphor of a Triple Helix more or less invites proposals to extend the model to more than three helices. In response to a discussion which focused on bringing "society" or "the public" back into the model as a fourth helix, Leydesdorff & Etzkowitz (2003) argued that the helices represent specialization and codification in function systems which evolve from and within civil society. A pluriform "society" is no longer coordinated by a central instance (such as "Rome" or "Moscow") but functions in terms of interactions among variously coded communications. These interactions can be expected to be both sources of variance (as in face-to-face communications) and "structurated" (Giddens, 1979) in terms of the different "horizons of meaning" (Husserl, 1929) being relevant for the interacting agents and institutions.[2]

Husserl's (1929) terminology of (potentially different) "horizons of meaning" can be elaborated sociologically into Parsons' (1963a, 1963b, 1968) and Luhmann's (1975, 1995) concept of "symbolically generalized media of communication." Money, for example, can be considered as a prime example of a symbolically generalized medium of communication: it enables us to pay without having to negotiate the price of a commodity. Power, truth, trust, and affection are other such media. While these media were confined in Parsons' oeuvre to his four meta-biological functions (adaptation, goal attainment, integration, and latency), Luhmann (following Merton) historicized the notion of possible functionalities in social communication.

---

[2] I prefer to use Giddens' (1979) term of "structuration" in this context because "structure" can be analyzed in networks of relations at each moment of time, for example, by using multivariate analysis (Leydesdorff, 2010a).

For example, one can raise the question of whether a new code has emerged at the interface between the sciences and the economy since patents became increasingly organized at the interfaces as a vehicle for the protection of intellectual property rights (Leydesdorff, 1996, 2008). Herbert Simon (1969, 1973) conjectured that any complex system operates with an alphabet. Thus, there may be 20+ symbolically generalizable media of communication available in interhuman interactions. While this plurality of codes can be expected to resound latently in inter-human interactions, some of the codes of communication can be specifically deselected in institutional settings.

For example, a subjective preference (or affection) is not considered as a legitimate argument in scholarly discourse. In court, one cannot offer the judge a payment in exchange for altering the verdict without corrupting the judicial system. In other words, the institutions which structure the expectations and communications that are considered legitimate in one institutional setting may be transgressive in another. This "constraining and enabling" (Giddens, 1979) takes place at each moment of time, whereas the communications and their codes develop over time.

Luhmann's (1995) model can be reformulated in terms of organized *integration* versus the *differentiation* that can be expected to prevail in the self-organization of the communication of meaning along the (analytically orthogonal) axes of the complex system (Achterbergh & Vriens, 2009). These two dimensions (integration because of organization at each moment of time and differentiation among the codes of communication over time) can then be considered as the woof and warp of social development. Innovative communications operate as variation in both directions, that is, by exploring new possibilities in a phase space of potentials—scientific, economic, political—and by recombining in instantiations here and now. Organization can thus be considered as a historically integrating function of this evolving system which also contains self-organizing dynamics.

From this perspective, the historicity of the development of the function of "organization" can also be specified (Leydesdorff, 2006, pp. 139 ff.). When organization prevails in the communication of meaning a stratified system can be expected, because organization operates in terms of relations. As noted, a hierarchy is necessarily shaped when relations are related recursively. At the level of society, such a stratified system can be considered as a high culture containing one "cosmological" order or another.

The disruption of this order as a consequence of the Reformation and Enlightenment—processes which took centuries—led to the installment of a "constructed" order following the invention of the *trias politica* in the 18$^{th}$ century. The "pursuit of happiness" as first codified in the American Constitution notably provided communication with two degrees of freedom—political and economic—and accordingly, political economies were developed into nation states during the 19$^{th}$ and 20$^{th}$ centuries. Academic freedom was yet not organized as a structural mechanism of change at the level of society, but considered as a prerogative institutionally guaranteed at the level of individual agency and provided as a shield to academic institutions.

The emergence of the alternative of a knowledge-based economy can be considered as a long-term historical process. German patent legislation of 1870 (Van den Belt & Rip, 1987), the so-called techno-scientific revolution which followed (1870-1910) after the stabilization of the major industrial powers during the 1860s and 1870s (Braverman, 1974; Noble, 1977), the scientification of warfare culminating in the Manhattan project during W.W. II, and the institutionalization of science & technology policies in the OECD countries during the 1970s and 1980s all prepared the ground for this gradual, but from the perspective of hindsight more fundamental transition.

The transition from a political economy to a knowledge-based economy became a major driver of the competition at the macro-level after the fall of the Berlin Wall and the demise of the Soviet Union (1990-1991). The political economy was gradually transformed into a knowledge-based economy because the battle between different ways of shaping political economies had become obsolete. The most explicit reflection of this transition was perhaps provided by the opening of China. Although different from the liberal model, the Chinese model remained flavored with evolutionary theorizing in terms of tensions and dialectics (Leydesdorff & Zeng, 2001).

**Conclusion**

I do not claim any *ex ante* or necessary limitation to three helices for the explanation of complex developments, but instead propose that an *N*-tuple or an alphabet of (20+) helices can be envisioned. However, in scholarly discourse and for methodological reasons, one may wish to extend models step-by-step and as needed to gain explanatory power. In the case of Japan, the addition of a fourth helix to the model was needed because along with university-industry-government relations internationalization also played an important role during the 1990s, that is, in reaction to the opening of China and the demise of the Soviet Union. With globalization, one can expect the international-national dimension to be increasingly relevant (Wagner, 2008). Globalization was further reinforced by the emergence of the internet as a medium beyond professional (academic and military) communication since the mid-1990s.

In other studies, the dimension of private versus public may be considered as yet another extension of the model. For the empirical researcher, much depends also on the availability of relevant data. Leydesdorff & Sun (2009) provide a model and rules for the calculation of triple-helix or more complex configurations in the study of Japan: the Triple Helix indicator—mutual information in three or more dimensions—changes sign with each selection mechanism added (Krippendorff, 2009; Leydesdorff, 2010). This is a consequence of the negative sign in a selection mechanism (Dolfsma & Leydesdorff, 2009). The theoretical task, however, remains the specification of each selection mechanism (Andersen, 1994).

Another advantage of using the Triple Helix model in qualitative research may be the increased awareness that the analysis of knowledge-based developments requires *at least* three relevant dimensions. For example, when the OECD analyzes regions (e.g.,

Piedmont; OECD, 2009), the administrative borders of such regions are taken for granted and thus the analysis is *a priori* reduced to the political economy of the region. Knowledge (e.g., patents) are considered as exogenous sources of economic activity and are not analyzed further than contextually (for example, in terms of their numbers). However, the patent portfolios of Piedmont and Lombardy may be complementary and synergetic, and a redefinition of geographical boundaries might be advisable on the basis of an analysis of the relevant knowledge-based subdynamics. I mentioned above as another example the transformation of what was assumed to be a "national" innovation system in the case of Hungary, whose three regions require different developmental strategies.

The Triple Helix model encourages the researcher to reflect on more than two possible dynamics (markets and governance). In a research project, one may—for pragmatic reasons—nevertheless wish to consider one of these contexts as given, but the reasons for this reduction should be deliberate and explained in the argument. Such an explanation can be expected to enrich the semantics because at least the three selection mechanisms are relevant for the study of knowledge-based developments.

One may wish to move beyond three relevant selection environments, but also a fourth (Carayannis & Campbell, 2009) or fifth (Carayannis & Campbell, 2010) dimension would require substantive specification, operationalization in terms of potentially relevant data, and sometimes the further development of relevant indicators. Without such a perspective, parsimony itself may be a methodologically well-advised strategy: so long as one is not able to operationalize and show development in the relatively simple case of three dimensions, one should be cautious in generalizing beyond the TH model to an *N*-tuple of helices.